\documentstyle[12pt,aaspp4]{article}

\begin{document}

\title{LARGE BODIES IN THE KUIPER BELT}
\author{Chadwick A. Trujillo\altaffilmark{1}}
\affil{Institute for Astronomy, 2680 Woodlawn Drive, Honolulu, HI
96822 \\ chad@ifa.hawaii.edu}
\author{Jane X. Luu}
\affil{Leiden Observatory, PO Box 9513, 2300 RA Leiden, The
Netherlands \\ luu@strw.leidenuniv.nl}
\author{Amanda S. Bosh}
\affil{Lowell Observatory, 1400 W. Mars Hill Road, Flagstaff, AZ
86001-4499 \\ amanda@lowell.edu}
\and
\author{James L. Elliot}
\affil{Department of Earth, Atmospheric, and Planetary Sciences and
Department of Physics, Massachusetts Institute of Technology,
Cambridge, MA 02139; and Lowell Observatory, Flagstaff, AZ 86001 \\
jle@mit.edu}
\altaffiltext{1}{Now at California Institute of Technology, MS 150-21,
Pasadena, CA 91125.  chad@gps.caltech.edu}

\slugcomment{Submitted to the {\it Astronomical Journal}}

\begin{abstract}
We present a survey for bright Kuiper Belt Objects (KBOs) and
Centaurs, conducted at the Kitt Peak National Observatory (KPNO) 0.9 m
telescope with the KPNO 8k Mosaic CCD.  The survey imaged 164 sq deg
near opposition to a limiting red magnitude of 21.1.  Three bright
KBOs and one Centaur were found, the brightest KBO having red
magnitude 19.7, about 700 km in diameter assuming a dark Centaur-like
4\% albedo.  We estimate the power-law differential size distribution
of the Classical KBOs to have index $q = 4.2^{+0.4}_{-0.3}$, with the
total number of Classical KBOs with diameters larger than 100 km equal
to $4.7^{+1.6}_{-1.0} \times 10^{4}$.  Additionally, we find that if
there is a maximum object size in the Kuiper Belt, it must be larger
than 1000 km in diameter.  By extending our model to larger size
bodies, we estimate that $30^{+16}_{-12}$ Charon-sized and
$3.2^{+2.8}_{-1.7}$ Pluto-sized Classical KBOs remain undiscovered.
\end{abstract}


\section{Introduction}

The region beyond Neptune is populated by $\sim 10^{5}$ Kuiper Belt
Objects (KBOs) with diameter $D > 100$ km, with a total mass of $\sim
0.2$ Earth masses (Jewitt, Luu, \& Trujillo 1998).  These bodies are
found in three dynamical classes: (1) the Classical KBOs, which have
semimajor axes in the $41 \mbox{ AU} < a < 47 \mbox{ AU}$ range with
low eccentricities, $e < 0.15$; (2) the Resonant KBOs, which occupy
the mean-motion resonances with Neptune, predominantly the Plutinos
($a \approx 39.4$ AU) and the 2:1 objects ($a \approx 47.7$ AU); and
(3) the Scattered KBOs, which are distant ($a > 50 \mbox{ AU}$) and
have highly eccentric orbits ($e \sim 0.6$; Jewitt \& Luu 2000).  To
date, very little is known about the physical properties of these
numerous Kuiper Belt Objects (KBOs) because most of the known objects
are quite faint; the known KBOs have median red magnitude $m_{R}
\approx 23.0$.  Studies of other outer Solar System objects suggest
that bright ($m_{R} < 20$) bodies are ideal targets for physical
studies (Brown et al. 1997; Luu \& Jewitt 1998; Brown, Cruikshank \&
Pendleton 1999; Luu, Jewitt \& Trujillo 2000; Brown 2000).  They may
also allow (1) albedo measurements through the use of combined thermal
and visible measurements, as collected from Centaur 10199 Chariklo
(provisional designation 1997 $\rm CU_{26}$; Jewitt \& Kalas 1998;
Altenhoff, Menten \& Bertoldi 2001) and KBO 2000 $\rm WR_{106}$
(Jewitt \& Aussel 2001); and (2) direct imaging with the Hubble Space
Telescope, such as performed on Pluto (Stern, Buie \& Trafton 1997)
and Centaur 2060 Chiron (Parker et al. 1997).  Large KBOs offer higher
chances for occultations and longer duration events, such as have been
observed for Chiron (Elliot et al. 1995; Bus et al. 1996), Charon
(Walker 1980; Elliot \& Young 1991), and Pluto (Elliot et al. 1989;
Millis et al. 1993).

Very few surveys have been sensitive to these bright KBOs, primarily
due to the fact that the surface density of ecliptic KBOs at $m_{R}
\sim 20$ is low, about 1 per 100 sq deg (Ferrin et al. 2001; Larsen et
al. 2001; and Trujillo, Jewitt \& Luu, 2001).  In the past,
photographic plate surveys have been used to search for these objects.
Tombaugh (1961) examined about 20,000 sq deg in a decade-long survey
to visual magnitude $m_{V} \sim 15.5$, and found only Pluto in 1930.
Kowal (1989) searched 6400 sq deg to visual magnitude $m_{V} \sim 20$
and found one Centaur, 2060 Chiron, but no KBOs.  Although plates
easily image large areas, they suffer from poor sensitivity and
surface defects.  In addition, computer-assisted analysis of plates is
more difficult than that of charge-coupled devices (CCDs).  More
recently, CCDs, which have a much higher quantum efficiency than
photographic plates, have played a growing role, due to technological
advancements allowing increased detector size.  For a review of all
published wide-field ($> 50$ sq deg) CCD surveys, see
Table~\ref{surveys}.

In this work we report results from a new survey of the outer solar
system undertaken at the Kitt Peak National Observatory (KPNO) using a
large format Mosaic CCD array (Muller et al. 1998).  We have surveyed
164 square degrees to limiting red magnitude $m_{R50} = 21.1$ (defined
as the magnitude at which survey efficiency drops to 1/2 the maximum
value).  This survey yielded 3 bright KBOs and 1 Centaur, the
brightest KBO being $m_{R} = 19.7$, corresponding to a diameter of 710
km, assuming a 4\% red albedo.  We use these data to constrain the number of
large KBOs and place limits on the maximum size of the KBOs using a
realistic model of the discovery process.

\section{Survey Data}
\label{survey}

\begin{figure}
\plotfiddle{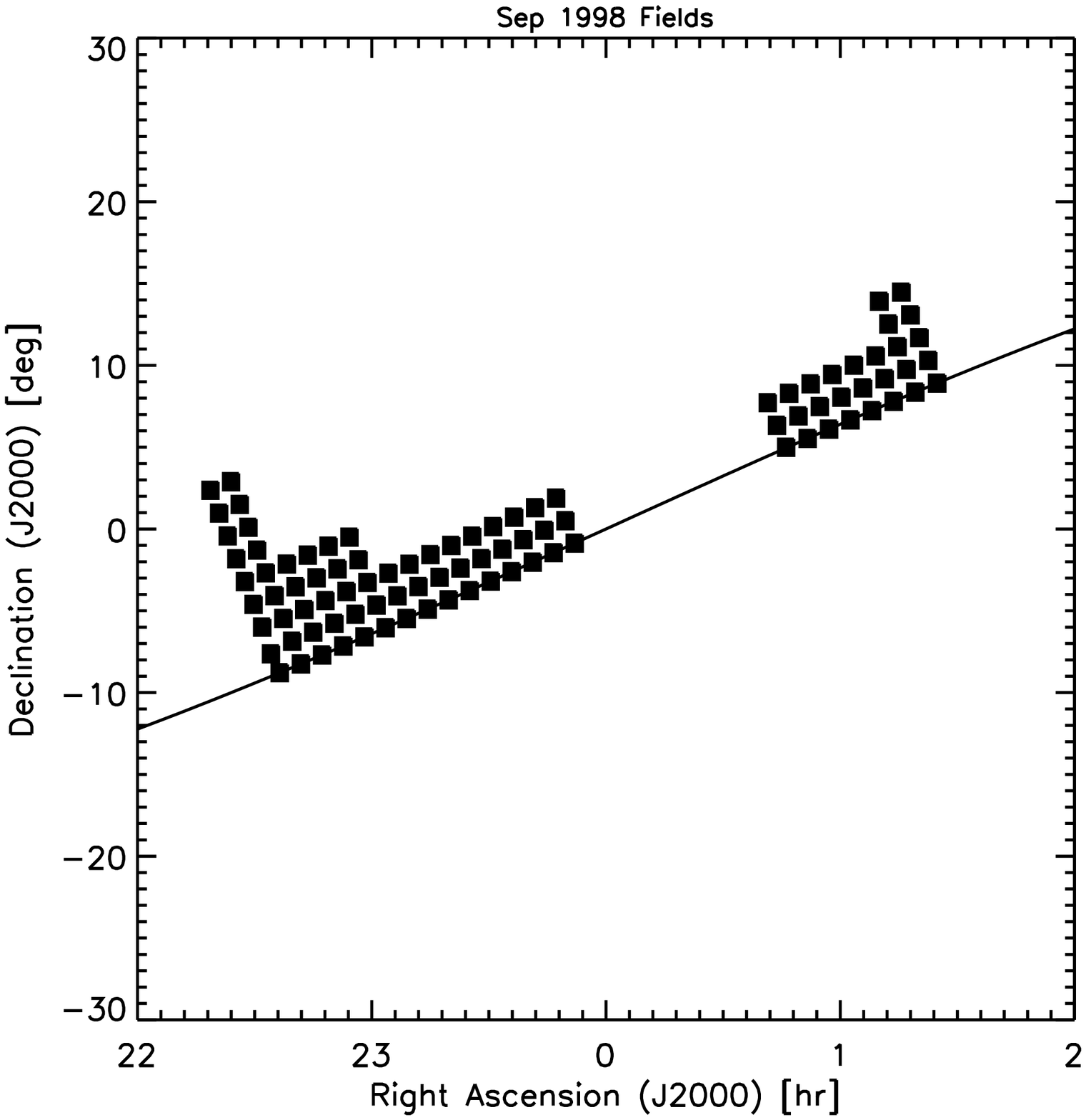}{6in}{0}{80}{80}{-250}{-100}
\figcaption{Fields that passed quality tests, imaged during Sep
1998. The ecliptic is denoted by a solid line. \label{fields2}}
\end{figure}

\begin{figure}
\plotfiddle{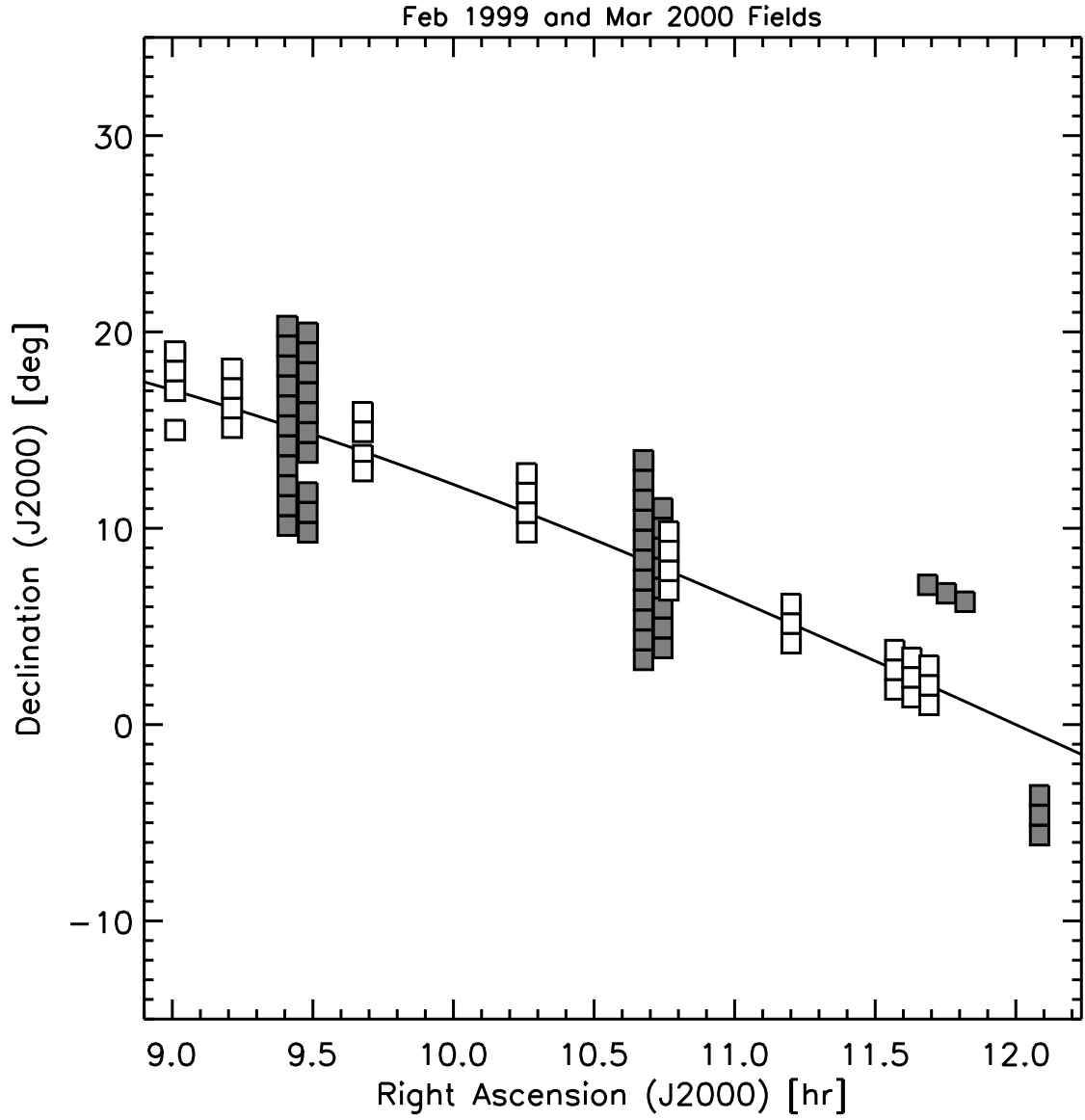}{6in}{0}{80}{80}{-250}{-100}
\figcaption{Fields that passed quality tests, imaged during Feb 1999
(white squares) and Mar 2000 (grey squares). The ecliptic is denoted
by a solid line. \label{fields1}}
\end{figure}

Observations were made with the now decommissioned KPNO 0.9 m f/7.5
telescope and the 8192 x 8192 pixel Mosaic CCD array.  This array is
composed of eight $4096 \times 2048$ pixel CCDs, each with an
independent bias level and flat-field response.  Observations were
made through a standard Johnson $R$ filter to enhance the detection of
the KBOs, which have $V-R \sim 0.6$.  The plate scale was 0.43
\arcsec/pixel for the 15 $\mu$m pixels, corresponding to a 0.96 sq deg
field of view.  This telescope and camera combination was unique in
providing a nearly 1 sq deg field of view while simultaneously
Nyquist-sampling stellar images.  In addition, the chips were
relatively free from defects and had typical quantum efficiencies of
$\sim 0.85$ in $R$.  Telescope and detector parameters are listed in
Table~\ref{kpnoobs}.  Fields were chosen to be within 1.5 hours of
opposition where the parallactic motion of the KBOs is greatest ($\sim
3$ \arcsec/hr), and easily distinguishable from the main-belt asteroid
motion ($\gtrsim 25$ \arcsec/hr).  Nearly all ($\sim$ 90\%) of the
fields were confined to be within 5 deg of the ecliptic, as depicted
in Figures \ref{fields2} and \ref{fields1}.  Each field was imaged
three times, using 200 s exposures with a $\sim 1$ hour separation
between images.  The observed fields included in the survey passed two
quality tests: (1) they must have been taken during photometric
conditions, and (2) they must have had a characteristic stellar
Full-Width at Half-Maximum (FWHM) better than 2.5 \arcsec.  Of the 285
fields imaged in the survey, only 171 passed these criteria.  These
fields are listed in Table~\ref{kpnofieldtable}.  As detailed in
Table~\ref{kpnoobs}, 1998 $\rm SN_{165}$ was discovered by Spacewatch
three nights before we serendipitously detected the object.  We
verified that no known objects remained undiscovered in our data by
computing the ephemerides of all bright ($m_{V} < 22.5$) objects with
opposition motion slower than 10 arcsec/hr for the epochs observed.

\begin{figure}
\plotfiddle{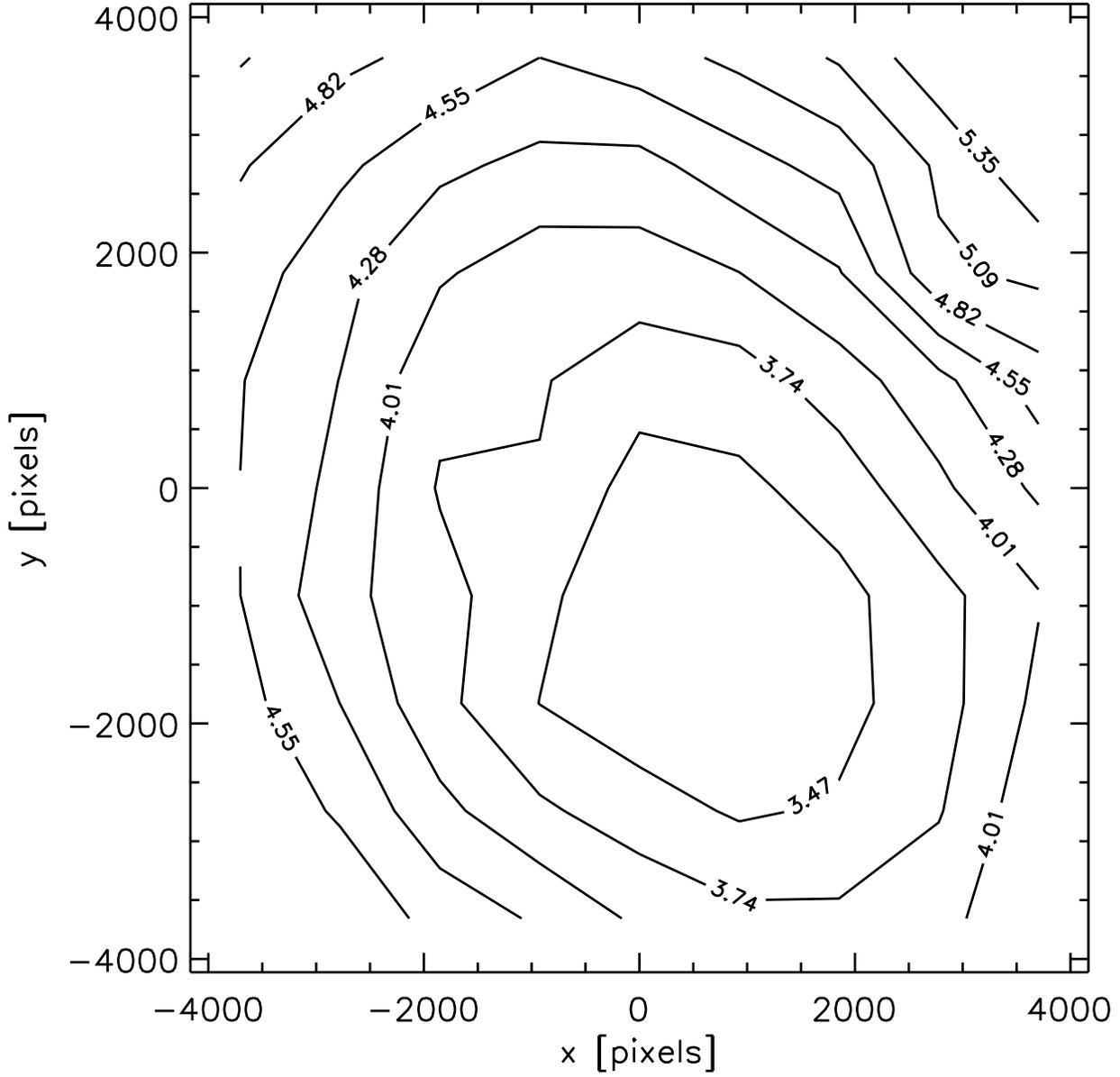}{7in}{0}{90}{90}{-300}{0}
\figcaption{Minor axis Moffat FWHM in pixels (1 pixel = 0.43 \arcsec)
of a stellar profile across the Mosaic CCD.  Note that the chip is
misaligned with the optical axis of the telescope by about 1500 pixels
= 2.3 cm.\label{psf}}
\end{figure}

The width, ellipticity, and position angle of the Point Spread
Function (PSF) varied significantly over the Mosaic CCD field of view.
We quantified this variation by fitting an elliptical Moffat PSF
(Moffat 1969) to stars in the linear flux regime of the CCDs for a
large number of images.  Figure~\ref{psf} shows the variations in the
FWHM of the PSF minor axis, as measured by IRAF's IMEXAM procedure.
The ellipticity was found to be correlated with the minor axis FWHM,
ranging in magnitude from 0.1 -- 0.3 over the focal plane.  The width,
ellipticity and position angle information was used to add simulated
KBOs to the data in order to test our Moving Object Detection Software
(MODS, Trujillo \& Jewitt 1998).

\begin{figure}
\plotfiddle{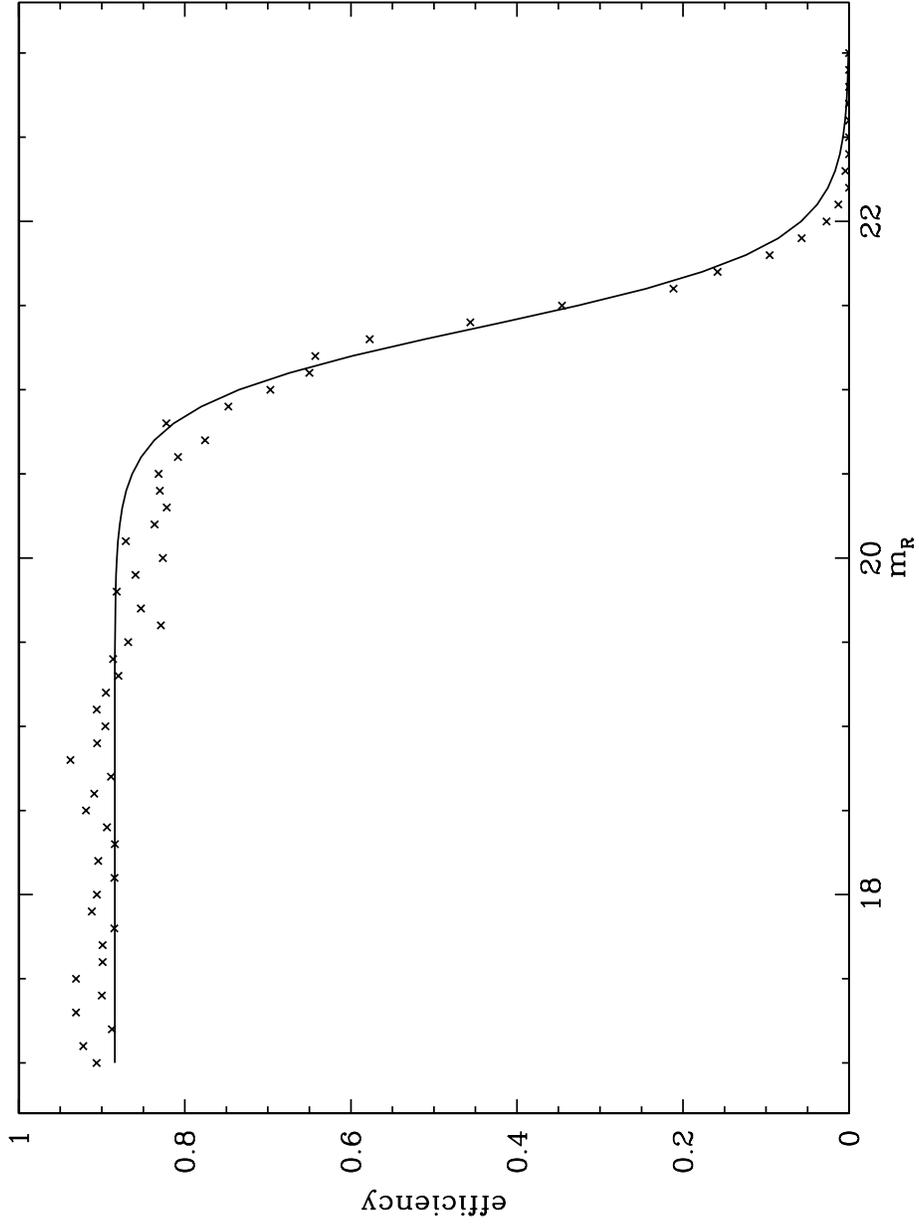}{7in}{0}{65}{65}{-200}{0}
\figcaption{The efficiency function of the 8k CCD on the KPNO 0.9m in
1.3\arcsec seeing, with a fit to the hyperbolic tangent efficiency
function. \label{kpnoefffig}}
\end{figure}

The seeing in the survey varied from 0.9\arcsec to 2.4\arcsec , with
median 1.5\arcsec.  We tested MODS with 3 groups of images based on
the seeing (0.9\arcsec -- 1.4\arcsec , 1.4\arcsec -- 1.9\arcsec , and
1.9\arcsec -- 2.4\arcsec ; with median seeing 1.3\arcsec , 1.6\arcsec
, and 2.1 \arcsec , respectively).  Artificial moving objects, with
profiles matched to the characteristic PSF for each image group and
location in the focal plane, were added to flattened images.  A tally
of objects was recorded, and the detection efficiency
(Figure~\ref{kpnoefffig}) was computed and found to be uniform with
respect to sky-plane speed in the 1.5 -- 10 \arcsec/hr range.

The PSF- and seeing-corrected efficiency function was fitted with the
same hyperbolic tangent function used in Trujillo, Jewitt \& Luu
(2000), given by
\begin{equation}
\label{kpnoeff}
\varepsilon =  \frac{\varepsilon_{\rm max}}{2} \left( \tanh \left( \frac{m_{R50} -
m_{R}}{\sigma} \right) + 1 \right).
\end{equation}
The limiting red magnitude $m_{R50}$ is the brightness where
$\varepsilon = \varepsilon_{\rm max}/2$, with $\varepsilon_{\rm max}$
defined as the maximum detection efficiency obtained for bright
objects.  The characteristic magnitude range over which the efficiency
drops from $\varepsilon_{\rm max}$ to zero is $\sigma$.  The values of
these parameters for each seeing range are listed in
Table~\ref{efftable}.  The poorer seeing images show a drop in
$m_{R50}$ compared to those with good seeing, consistent with reduced
signal-to-noise ratio.  Thus, the limiting magnitude can be
characterized as $m_{R50} = 21.8 - 2.5 \log (\theta)$, where $\theta$
represents the FWHM of a stellar image in arc sec.  The quantity
$\varepsilon_{\rm max}$ varies little for the different seeing
categories; even for the worst seeing, $\varepsilon_{\rm max}$ is
reduced by only 9\% from the best seeing.

The discovery circumstances for the 4 KPNO objects appear in
Table~\ref{kpnodisc}.  Although the discovered objects were found
during good seeing (Table~\ref{kpnoobs}), the null detection of
objects in the medium and poor seeing cases is not statistically
significant.  Given the $\alpha = 0.66$ slope of the Cumulative
Luminosity Function (\S~\ref{results}) and the limiting magnitude and
sky area presented in Table~\ref{efftable}, 46\% of the objects should
be found during good seeing.  Thus, the probability of all 4 object
detections occurring by chance during good seeing is $0.46^{4} = 4.4\%$
(less probable than a $2 \sigma$ Gaussian event).  We therefore
combine the data for all seeing cases into a single global efficiency
function.

\section{The Kuiper Belt}
\label{results}

The four critical quantities we estimate for the KBOs are: (1) the
Cumulative Luminosity Function, or CLF, (2) the size distribution and
total number of objects, (3) the KBO maximum size, and (4) the total
number and mass of large KBOs ($\gtrsim 2000$ km and $\gtrsim 1000$ km
diameter).

\subsection{The Cumulative Luminosity Function}
\label{clfsection}

The Cumulative Luminosity Function (CLF) describes the surface density
of KBOs near the ecliptic ($\Sigma$) versus limiting red survey
magnitude ($m_{R}$) and is fit by the equation
\begin{equation}
\log \Sigma = \alpha(m_{R} - m_{0}),
\end{equation}
where $\alpha$ describes the slope and $m_{0}$ is the red magnitude at
which $\Sigma = 1$ KBO per sq deg.  Assuming heliocentric and
geocentric distance and albedo independent of object size, and a
differential size distribution $n(r) dr \propto r^{-q} dr$, where
$n(r) dr$ describes the number of bodies with radii between $r$ and
$r+dr$, the size distribution of the KBOs can be directly measured via
(Irwin et al. 1995)
\begin{equation}
\label{qalpha}
q = 5 \alpha + 1 .
\end{equation}

We estimated the CLF by combining the discovery results of the 86 KBOs
found by Trujillo, Jewitt \& Luu (2001, hereafter 01TJL) with the
three KBOs found in our data.  Although the 01TJL data dominate the
total numbers of faint objects, the two surveys have found comparable
numbers of bright objects.  In addition, our brightest object (1999
$\rm DE_{9}$, $m_{R} = 19.7$) was over a magnitude brighter than the
brightest object from 01TJL (1999 $\rm CD_{158}$, $m_{R} = 21.0$).
These two data sets improve upon previous survey results due to
discovery statistics that cover nearly a 5 magnitude range ($19.7 <
m_{R} < 24.4$).

\begin{figure}
\plotfiddle{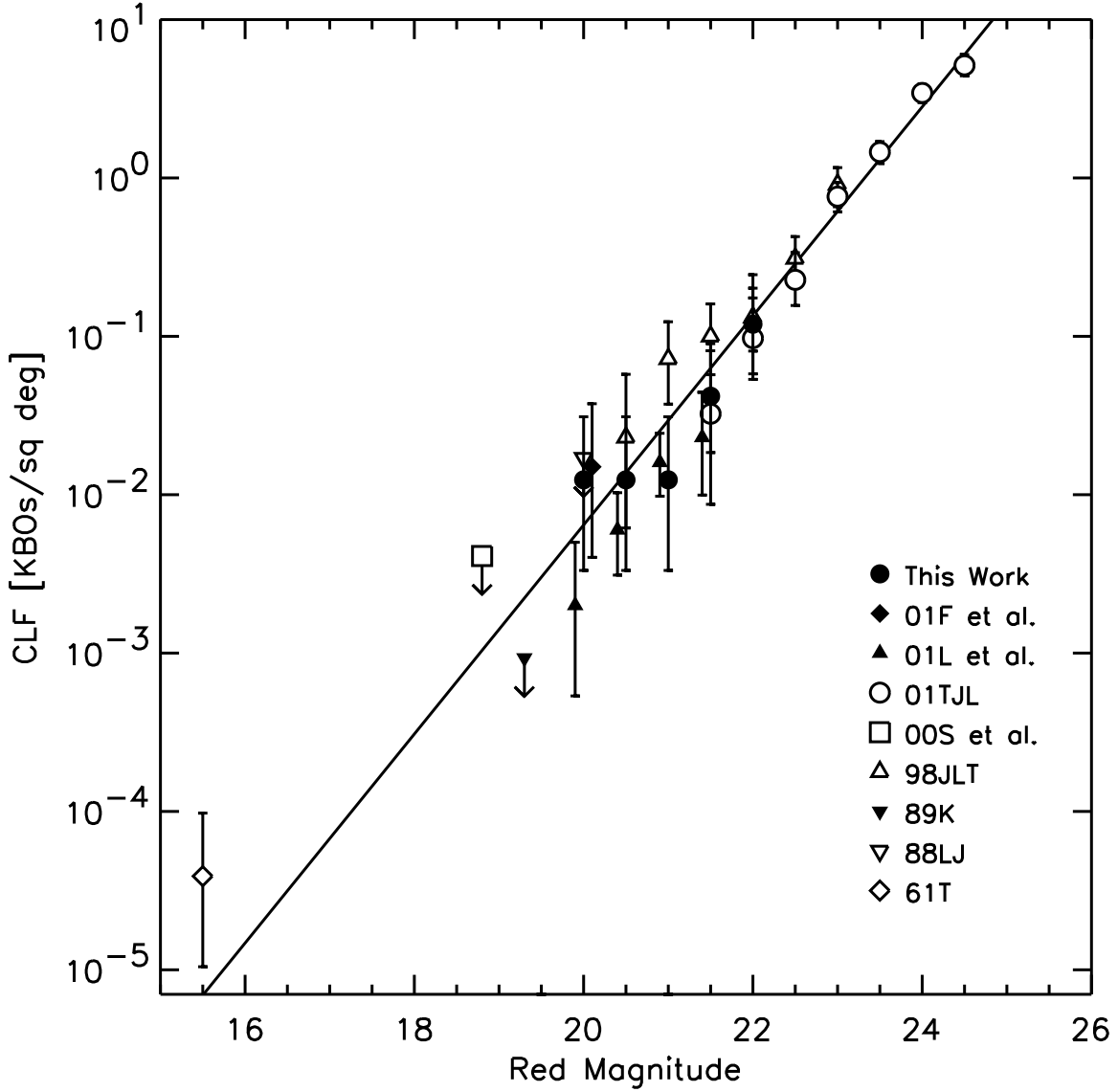}{6in}{0}{80}{80}{-250}{-100}
\figcaption{Our measurement of the Cumulative Luminosity Function
(CLF), which represents the number of KBOs $\mbox{deg}^{-2}$ near the
ecliptic (filled circles) brighter than a given apparent red
magnitude.  The line represents the best fit ($\alpha = 0.66 \pm 0.06$
and $m_{0} = 23.32 \pm 0.09$) to the CLF determined by this
survey and that conducted by Trujillo, Luu \& Jewitt (2001, 01TJL).
Other points are from previous works: 01F et al. (Ferrin et al. 2001);
01L et al. (Larsen et al. 2001); 00S et al. (Sheppard et al. 2000);
98JLT (Jewitt, Luu \& Trujillo 1998); 89K (Kowal 1989); 88LJ (Luu \&
Jewitt 1988); and 61T (Tombaugh 1961), with arrows denoting upper
limits.  \label{clf}}
\end{figure}

Since the CLF is a cumulative measure, the CLF value at a given
magnitude is correlated with the CLF values at fainter magnitudes.
Thus, a linear fit to the CLF will implicitly weight the bright bodies
(which are counted in the bright and faint data points) to a greater
degree than the faint bodies (which are only counted in the faint data
points).  We instead perform a maximum-likelihood fit to find $\alpha$
and $m_{0}$ (Gladman et al. 1998), using the efficiency function
(Equation~\ref{kpnoeff}) and the sky area imaged (164 sq deg).  It was
impossible for all fields to be centered on the ecliptic due to the
large sky area imaged.  Thus, sky coverage was computed by weighting
each field imaged by the fraction of KBOs expected at that ecliptic
latitude, assuming the best-fit inclination distribution of 01TJL, a
half-width of 20 deg.  For our KPNO data, this correction reduces our
164 sq deg to an effective area of 95.7 sq deg, resulting in a minor
change (23\%) in the normalization of $\log(\Sigma)$, a factor 2
smaller than the $1 \sigma$ Poisson uncertainty for $\log(\Sigma)$ due
to the 4 KBOs discovered (45\%).  The resulting best fit follows
$\alpha = 0.66 \pm 0.06$ and $m_{0} = 23.32 \pm 0.09$.  This result is
in statistical agreement with many other previous works (Jewitt, Luu
\& Trujillo 1998 with $\alpha = 0.58 \pm 0.05$, $m_{0} = 23.27 \pm
0.11$; Gladman et al. 1998 with $\alpha = 0.76^{+0.10}_{-0.11}$ and
$m_{0} = 23.40^{+0.20}_{-0.18}$; Chiang \& Brown 1999 with $\alpha =
0.52 \pm 0.05$ and $m_{0} = 23.5 \pm 0.06$; and 01TJL $\alpha = 0.63
\pm 0.06$, $m_{0} = 23.04^{+0.08}_{-0.09}$).  We present this simple
fit, our observations, and other works covering $> 50$ sq deg in
Figure~\ref{clf}.  Using equation~\ref{qalpha}, we find the exponent
of the differential size distribution to be $q = 4.3 \pm 0.3$.  We
present a more detailed analysis in the next section, where we use a
more realistic simulation to constrain the slope of the size
distribution.

\subsection{The Size Distribution of the KBOs}

To better constrain the size distribution, we estimate $q$ and the
total number of Classical KBOs larger than 100 km, $N_{\rm CKBOs}(D >
100 \mbox{ km})$, using a more realistic maximum-likelihood simulation
than the CLF fit described in the previous section.  Bias factors such
as the detection efficiency and the distribution of discovery
distances are correctly handled in this more detailed simulation.  A
full description of our simulation can be found in Trujillo, Jewitt \&
Luu (2000) and 01TJL.  We summarize the assumed quantities in
Table~\ref{model}, and key procedures here:

(1) Simulated orbital elements and object sizes were drawn for the
    Classical KBOs (CKBOs).

(2) Object brightnesses were computed from the following formula
    (Jewitt \& Luu 1995): \begin{equation} \label{mreq} m = m_{\odot}
    - 2.5 \log(p \Phi(\alpha') r^{2}) + 2.5 \log(2.25 \times 10^{16}
    R^{2} \Delta^{2}), \end{equation} where $\alpha'$ represents the
    phase angle of the observations, $\Phi(\alpha')$ is the Bowell et
    al. (1989) phase function for dark bodies ($G = 0.15$), geometric
    albedo is given by $p$, object radius is described by $r$ [km],
    and $R$ [AU] and $\Delta$ [AU] describe the heliocentric and
    geocentric distance of the body, respectively.  The apparent red
    magnitude of the Sun is $m_{\odot} = -27.1$.

(3) Object ecliptic coordinates and velocities were computed from the
    equations of Sykes \& Moynihan (1996, sign error corrected).

(4) The observed field areas and efficiency functions were used to
    determine which of the simulated objects could be detected in
    either our survey or that of 01TJL, based on the ecliptic
    coordinates, velocities and brightnesses computed in steps 2 and
    3.

(5) Two tallies of the radii of the ``detected'' simulated objects
    were kept, one for our KPNO 8k Mosaic survey and one for the 01TJL
    survey.  Steps 1-4 were repeated until there were a factor 10 to
    100 more simulated objects in each radius bin than were actually
    observed.

(6) The likelihood of the model producing the observed distribution of
    radii was estimated by assuming Poisson detection statistics with
    the number of ``detected'' simulated objects providing the
    expectation value for each bin.  This dataset consists of the 57
    CKBOs found by 01TJL, and the single CKBO found in the KPNO survey
    (1999 $\rm DF_{9}$).  Although the 01TJL data contains many more
    objects, the KPNO survey has comparable area coverage, and number
    of bright object discoveries.

(7) Steps 1-6 were repeated, each time varying the size distribution
    index, $q$, and the total number of Classical KBOs, $N_{\rm
    CKBOs}(D > 100 \mbox{ km})$, with the ultimate goal of finding the
    number of CKBOs and size distribution that has the maximum
    likelihood of producing the observations.

The primary process we are modeling is the effect of heliocentric
distance $R$ and geocentric distance $\Delta$ on the apparent red
magnitude $m_{R}$, and the resulting probability of detection based on
detection efficiency (Equation~\ref{kpnoeff}) and survey area.  The
inclination distribution assumed in Table~\ref{model} is equivalent to
the best-fit value found by 01TJL --- a Gaussian model with half-width
of 20 deg.  The true ecliptic latitude distribution of observed fields
was used in the simulation, so no explicit correction for finite belt
thickness is necessary, as needed in the CLF fit
(\S~\ref{clfsection}).

\begin{figure}
\plotfiddle{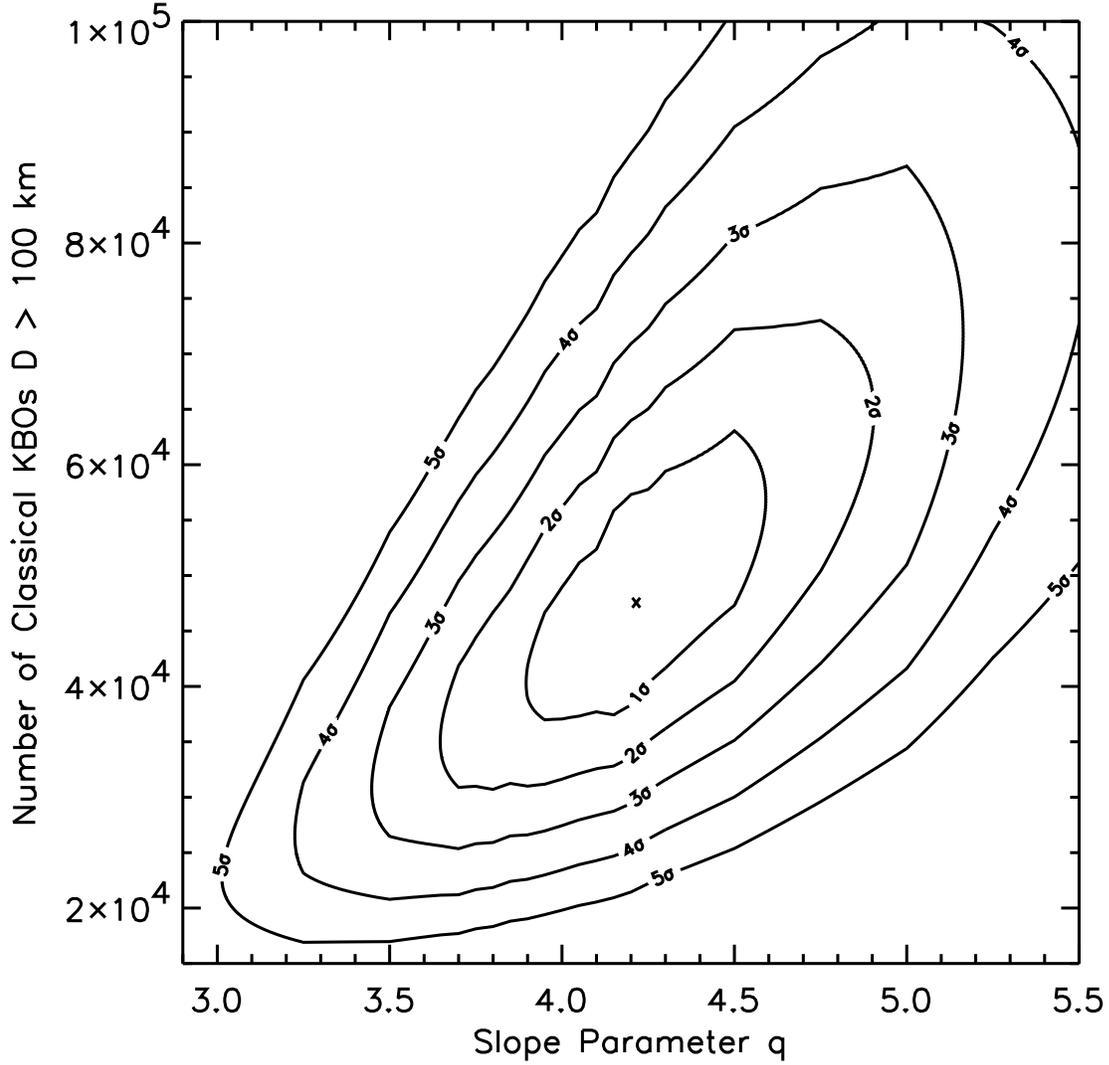}{6in}{0}{80}{80}{-250}{-100} \figcaption{The
normalized probability that the observed distribution of radii match
the expected distribution given the free parameters: (1) $q$, the size
distribution index, and (2) $N_{\rm CKBOs}(D > 100 \mbox{ km})$, the
total number of Classical KBOs (sigma values correspond to Gaussian
confidence limits: $1 \sigma$ = 68.3\%, $2 \sigma$ = 95.4\%, etc.).
The maximum likelihood occurs at $q = 4.2^{+0.4}_{-0.3}$ ($1 \sigma$).
\label{kboqdist}}
\end{figure}

The results for the best-fit size distribution index, $q$, and number
of Classical KBOs with diameters $D > 100$ km, $N_{\rm CKBOs}(D > 100
\mbox{ km})$, are presented in Figure~\ref{kboqdist}.  This figure
depicts the contours of constant probability that the given $q$ and
$N_{\rm CKBOs}(D > 100 \mbox { km})$, could produce the observed
distribution. We find that
\begin{displaymath}
\begin{array}{rcll}
q        & = & 4.2^{+0.4}_{-0.3} & (1 \sigma) \mbox{ and} \\
         & = & 4.2^{+1.0}_{-0.8} & (3 \sigma), \\
         & \mbox{and} & & \\
N_{\rm CKBOs}(D > 100 \mbox{ km}) & = & 4.7^{+1.6}_{-1.0} \times 10^{4} & (1 \sigma) \mbox{ and} \\
                                  & = & 4.7^{+4.0}_{-2.2} \times 10^{4} & (3 \sigma). \\
\end{array}
\end{displaymath}
This estimate of $q$ is in formal agreement with that derived from the
CLF alone at $<1 \sigma$ level.

The results for $q$ to not depend sensitively on our model
assumptions.  Our values for $N_{\rm CKBOs}(D > 100 \mbox{ km})$ are
somewhat dependent on the inclination distribution assumed, as thicker
inclination distributions can hide more objects at high ecliptic
latitudes where no observations were made.  For example, increasing
the half-width of the inclination distribution by a factor of two,
from 20 deg to 40 deg results in an increase in the best-fit $N_{\rm
CKBOs}(D > 100 \mbox { km})$, from $4.7 \times 10^{4}$ to $7.5 \times
10^{4}$, a factor 1.6.  The best-fit value for $N_{\rm CKBOs}(D > 100
\mbox{ km})$ is roughly linear with half-width $i_{1/2}$, following
$N_{\rm CKBOs}(D > 100 \mbox{ km}) \sim (1.9 + 0.14 i_{1/2}) \times
10^{4}$.

\subsection{The Maximum Size of Kuiper Belt Objects}
\label{maxsizesection}

\begin{figure}
\plotfiddle{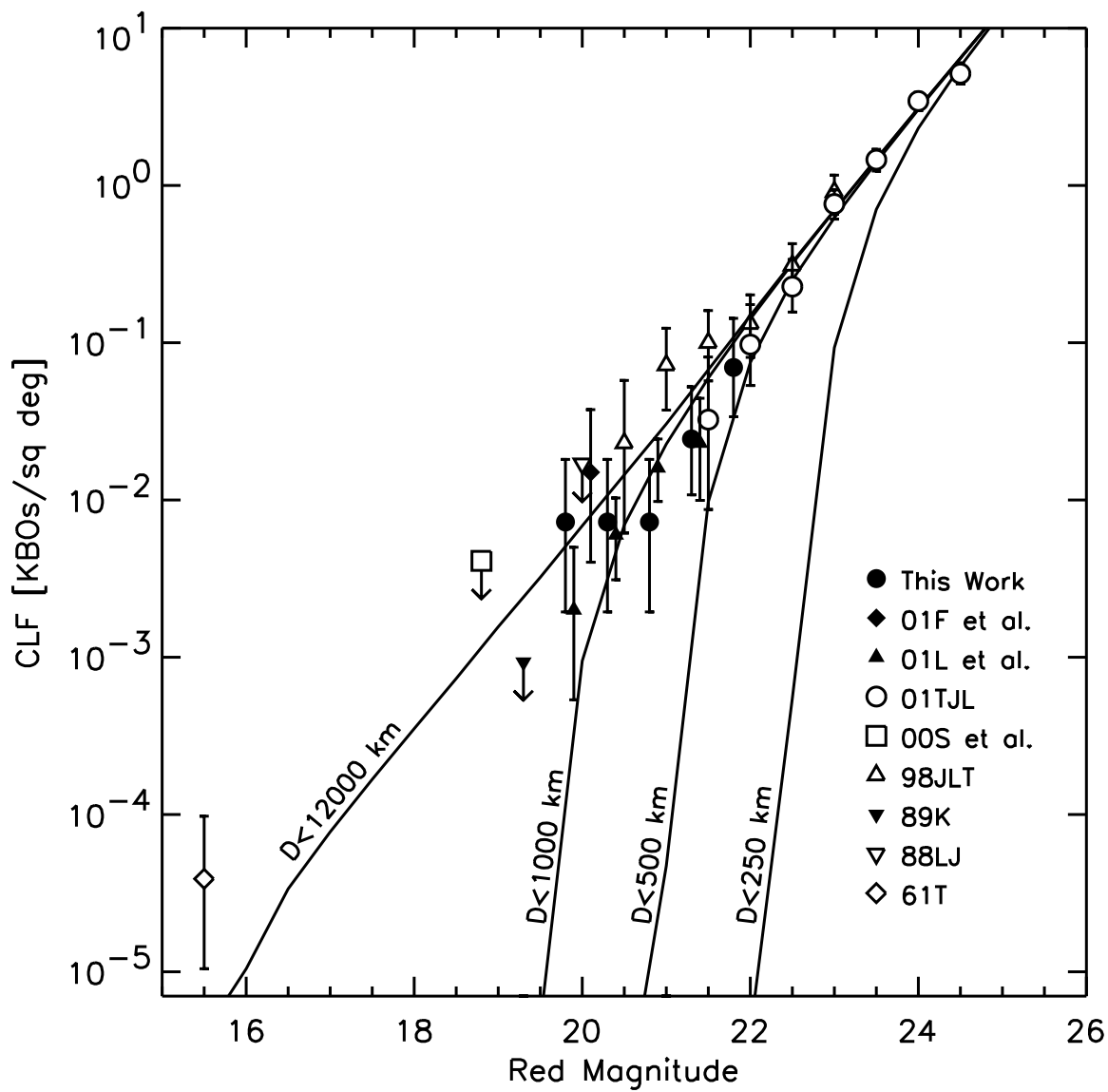}{6in}{0}{80}{80}{-250}{-100}
\figcaption{The expected observed CLF for the $q = 4.2$ case, given
maximum KBO diameters (left-to-right) of 12,000 km ($\sim$
Earth-sized), 1000 km ($\sim$ Pluto-sized), 500 km ($\sim$
Charon-sized) and 250 km.  The 1000 km case represents the best fit,
however only the 250 km and 500 km cases are rejected at the $3
\sigma$ level.  Formally, there is no evidence for an upper-size
cutoff.
\label{maxradius}}
\end{figure}

Adopting the best-fit $q=4.2$, we simulated the CLF that would be
found by our survey given several different maximum sizes for the
Classical KBOs, with results appearing in Figure~\ref{maxradius}.  The
expected CLF from the Classical KBOs was multiplied by the observed
KBO : Classical KBO ratio (89:58) to construct this graph, which makes
the implicit assumption that the CLF of the Classical KBOs matches
that of the KBOs as a whole.  This assumption is reasonable, as a
maximum-likelihood fit to the CLF for the Classical KBOs alone
($\alpha = 0.71^{+0.07}_{-0.06}$) is statistically consistent with the
fit to all observed populations ($\alpha = 0.66 \pm 0.06$,
\S~\ref{clfsection}).

Our analysis rules out the $D_{\rm max} = 500$ km and $D_{\rm max} =
250$ km diameter maximum sizes, as is expected since we found one body
with diameter $D \sim 700$ km and another with $D \sim 500$ km.  The
$D_{\rm max} = 1000$ km simulation is the best-fit to our data,
however, these data are formally consistent will all greater $D_{\rm
max}$.  Thus, if there is a maximum size cutoff, it is greater than or
equal to $D_{\rm max} = 1000$ km, because we do not observe a
significant surface density decrease for bright objects beyond that
described by the linear slope of the CLF.  This analysis is in
agreement with all other published wide-field surveys except for Kowal
(1989).  Kowal's (1989) datum could be explained by a precipitous drop
in surface density between $19 < m_{R} < 20$, as would be expected if
there were a maximum size of $D = 1000$ km to the KBOs.  Such a model
would be in violation of the Tombaugh (1961) datum point, unless Pluto
is considered a special case.  In either case, our observations
indicate that if there is a maximum size, it must be $D_{\rm max} \geq
1000 \mbox{ km}$.

\subsection{The Total Number and Mass of Large Kuiper Belt Objects}

The number of large KBOs is of fundamental interest because it is a
direct measurement of the runaway accretion in the planet forming era
of the solar system (Kenyon \& Luu 1998).  Two sizes of typical
interest are the number of Charon-sized bodies ($\sim 1000$ km in
diameter) and the number of Pluto-sized bodies ($\sim 2000$ km in
diameter).  The number of bodies with diameters larger than $D_{\rm
min}$, given by $N(D > D_{\rm min})$, can be directly calculated from
the results of our maximum likelihood simulation.  For $q \neq 1$,
\begin{equation}
N(D > D_{\rm min}) = N_{\rm CKBOs}(D > 100 \mbox{ km})
\left(\frac{100 \mbox{ km}}{D_{\rm min}} \right)^{q-1}.
\end{equation}

Thus, from Figure~\ref{kboqdist}, we find that the total numbers of
Charon- and Pluto-sized bodies in the Classical Kuiper Belt follow
\begin{displaymath}
\begin{array}{rcll}
N(D > 1000 \mbox{ km}) & = & 30^{+16}_{-12} & (1 \sigma) \mbox{ and} \\
                       & = & 30^{+57}_{-22} & (3 \sigma), \\
                       & \mbox{and} & & \\
N(D > 2000 \mbox{ km}) & = & 3.2^{+2.8}_{-1.7}   & (1 \sigma) \mbox{ and} \\
                       & = & 3.2^{+12.2}_{-2.7}  & (3 \sigma). \\
\end{array}
\end{displaymath}
Thus, assuming that the $q=4.2$ size distribution continues to large
sizes, we expect that there are $\geq 1$ ($3 \sigma$ lower-limit)
undiscovered Pluto-sized bodies in the Classical Kuiper Belt and $\geq
8$ ($3 \sigma$ lower-limit) Charon-sized bodies, of which only a few
have been discovered.  If the size distributions of the Classical and
Scattered KBOs are identical, then a similar number of bodies should
be present in the Scattered Kuiper Belt, which has approximately the
same number of bodies (Trujillo, Jewitt \& Luu 2000). These results
are consistent with the Kuiper Belt growth models of Kenyon and Luu
(1998, 1999) using velocity evolution and collisional fragmentation.
In these models, several Pluto-sized bodies grow concurrently in a
low-mass ($\sim 10 M_{\earth}$) disk.  These models have two
characteristic observable parameters: (1) The size distribution of the
resulting Kuiper Belt corresponds to $q \approx 4$, and (2) several
large bodies are formed concurrently.  Their favored model produces
$\sim 150$ Charon-sized bodies and 1 Pluto-sized body in 36.5 Myr
(Kenyon and Luu 1999a) and is roughly consistent with our
observations, although no prediction is made about the subsequent
destruction rates of such bodies from catastrophic collisions.

The possibility of finding these Pluto- and Charon- sized bodies can
be estimated by examining the discovery conditions of other bright
bodies.  Table~\ref{brightdisc} lists the three KBOs to date brighter
than red magnitude $m_{R} = 20$ that have been found in published
surveys (with the exception of Pluto).  These three bodies were found
in the heliocentric distance range 30--39 AU.  Assuming opposition
observations and 4\% albedo, 2000 km and 1000 km diameter bodies at
these heliocentric distances would have have brightnesses $16.9 <
m_{R} < 18.2$ and $18.4 < m_{R} < 19.7$, respectively, according to
Equation~\ref{mreq}.  These magnitudes are within range for most
time-resolved wide-field survey telescopes, however the apparent
motion of these bodies, $\sim 3$ \arcsec/hr, is considerably slower
than $\gtrsim 25$ \arcsec/hr main-belt asteroids.  Since most such
observing programs are designed to find near-Earth asteroids and other
high proper motion planetary bodies, plate scales and timebases
between images are not usually conducive to finding the KBOs.  Thus,
special observing procedures or software methods must be adopted to
find the slower KBOs, as has been done by Spacewatch (Larsen et
al. 2001).  A time-resolved all-sky $m_{R} = 20.0$ magnitude survey
should be able to find all the remaining Pluto- and Charon- sized
bodies, as long as suitable measures are taken to detect $\sim 3$
\arcsec/hr bodies.  Such a survey could be conducted with a dedicated
1 meter class telescope outfitted with a CCD camera allowing a
wide-field of view, similar to the KPNO 0.9 m and 8k CCD Mosaic
instrument combination used in this work.

These analyses presume that the KBOs have a 4\% albedo, consistent
with the Centaurs (Jewitt \& Luu 2000).  If the mean KBO albedo, $p$,
departs from this value, then the radial sizes of the bodies will be
affected by a factor $(p/0.04)^{-1/2}$.  For a 14\% albedo such as
that of Chiron (Campins et al. 1995) and a 38\% albedo such as found
for Charon (Buie, Tholen \& Wasserman 1997), this amounts to a size
factor 1.9 and 3.1, respectively.

The total amount of mass in 1000 km to 2000 km diameter Classical KBOs
for our best-fit $q = 4.2$ case is
\begin{equation}
M(1000 \mbox{ km} < D < 2000 \mbox{ km}) = 5.4 \times 10^{-3}
M_{\earth} \left(\frac{\rho}{1000 \mbox{ kg} \mbox{ m}^{-3}}\right)
\left(\frac{0.04}{p}\right)^{1.6},
\end{equation}
with $\rho$ representing the bulk density, $p$ denoting the albedo,
and $M_{\earth} = 6.0 \times 10^{24}$ kg.  This accounts for 18\% of
the total mass found in $100 \mbox{ km} < D < 2000 \mbox{ km}$ bodies.
Again, the Scattered KBOs may contain a similar amount of material.

\section{The Centaurs}

Although there is no formal International Astronomical Union
definition of the Centaur population, we classify bodies whose
perihelia fall between that of Jupiter and Neptune as Centaurs,
similar to the Larsen et al. (2001) definition.  One such Centaur was
found in the KPNO 8k data (2000 $\rm EE_{173}$).  We therefore
estimate the sky-plane density of Centaurs to be $1.0_{-0.8}^{+1.6}
\times 10^{-2}$ ($1 \sigma$) Centaurs per sq deg at our survey
limiting magnitude $m_{R50} = 21.1$.  This is consistent with other
recent estimates of the Centaur population: Larsen et al. (2001)
predict a value of $1 \times 10^{-2}$ at our limiting magnitude and
Sheppard et al. (2000) predict a similar surface density of $3 \times
10^{-3}$ Centaurs per sq deg.

\section{Summary}
We have conducted a survey for KBOs and Centaurs with the KPNO 0.9 m
telescope and 8k Mosaic CCD, which together provide a nearly 1 sq deg
field of view, reaching red magnitude $m_{R} = 21.1$ in 200 s in
typical seeing conditions at Kitt Peak.  We detected 3 KBOs and 1
Centaur in 164 sq deg examined near the ecliptic.  Combining these new
data with the results of Trujillo, Jewitt \& Luu (2001), we find the
following results:

(1) The Cumulative Luminosity Function slope is $\alpha = 0.66 \pm
    0.06$, with 1 KBO / sq deg occurring at red magnitude $m_{0} = 23.32
    \pm 0.09$.

(2) The Classical KBOs follow a $q=4.2^{+0.4}_{-0.3}$ ($1 \sigma$)
    differential size distribution, with the total number of Classical
    KBOs with diameters larger than 100 km given by $N_{\rm CKBOs}(D >
    100 \mbox{ km}) = 4.7^{+1.6}_{-1.0} \times 10^{4}$, assuming a
    Gaussian inclination distribution with half-width 20 deg.

(3) No statistically significant evidence for a maximum size
    cutoff to the KBOs was found.  However, if such a maximum diameter
    exists, it must have $D_{\rm max} > 1000$ km.

(4) Assuming the differential size distribution extends to large
    sizes, the total number of Classical KBOs with diameters $D
    \gtrsim 1000$ km and $D \gtrsim 2000$ km is $30^{+16}_{-12}$ and
    $3.2^{+2.8}_{-1.7}$, respectively ($1 \sigma$).  The bodies in the
    $1000 \mbox{ km} < D < 2000 \mbox{ km}$ diameter range account for
    $\sim 1/5$ of the total mass in the $100 \mbox{ km} < D < 2000
    \mbox{ km}$ diameter range, representing $M(1000 \mbox{ km} < D <
    2000 \mbox{ km}) = 5.4 \times 10^{-3} M_{\earth}$, assuming a bulk
    density of 1000 $\mbox{ kg} \mbox{ m}^{-3}$.

(5) The sky-plane surface density of Centaurs brighter than $m_{R} =
    21.1$ is $1.0_{-0.8}^{+1.6} \times 10^{-2}$ ($1 \sigma$)
    Centaurs/sq deg near the ecliptic.

\acknowledgements

We thank the National Optical Astronomy Observatories (NOAO) for
granting telescope time, David Tholen for computation of the orbital
elements, Amanda Sickafoose and Wyn Evans for observational
assistance, David Jewitt for editorial comments, and Nigel Sharp for
technical assistance at the 0.9m telescope.  This work was supported
in part by a NASA grant to David C. Jewitt, NASA grant NAG5-3940 to
JLE and through NSF support provided to NOAO.

%\section{Figure Captions}

\clearpage

\clearpage

\clearpage

\clearpage

\clearpage

\clearpage

\clearpage

\clearpage

\begin{center}
\begin{deluxetable}{rccrrl}
\small
\tablewidth{0pt}
\tablecaption{Wide-Field CCD Surveys sensitive to KBOs}
\tablehead{\colhead{Area} & \colhead{$m$\tablenotemark{a}} &
\colhead{Centaurs\tablenotemark{b}} & \colhead{KBOs\tablenotemark{c}}
& \colhead{Brightest KBO} & \colhead{Source} \\ \colhead{[sq deg]} &
\colhead{} & \colhead{} & \colhead{} & \colhead{} & \colhead{}}
\startdata
1428 & 18.8 $R$ &  0 (1) &  0 (0) & ---                                  & Sheppard et al. (2000) \nl
 522 & 20.5 $V$ &  3 (1) &  7 (5) & 1999 $\rm GQ_{21}$  ($m_{V} = 20.5$) & Larsen et al. (2001) \nl
 164 & 21.1 $R$ &  1 (0) &  2 (1) & 1999 $\rm DE_{9}$   ($m_{R} = 19.7$) & This Work \nl
  73 & 23.7 $R$ &  0 (0) & 84 (2) & 1999 $\rm CD_{158}$ ($m_{R} = 21.0$) & Trujillo, Jewitt \& Luu (2001) \nl
  67 & 20.1 $R$ &  0 (0) &  1 (0) & 2000 $\rm EB_{173}$ ($m_{R} = 19.3$) & Ferrin et al. (2001) \nl
  52 & 22.5 $R$ &  0 (0) & 13 (0) & 1996 $\rm TL_{66}$  ($m_{R} = 20.6$) & Jewitt, Luu \& Trujillo (1998) \nl
\enddata
\tablecomments{Published CCD surveys with $> 50$ sq deg have been included.}
\tablenotetext{a}{Limiting magnitude of survey.}
\tablenotetext{b}{The number of Centaurs discovered is listed, with
serendipitous detections of previously discovered objects listed in
parentheses.}
\tablenotetext{d}{The number of KBOs discovered is listed, with
serendipitous detections of previously discovered objects listed in
parentheses.}
\label{surveys}
\end{deluxetable}
\end{center}

\clearpage

\begin{center}
\begin{deluxetable}{cc}
\small
\tablewidth{0pt}
\tablecaption{KPNO Survey Parameters}
\tablehead{\colhead{Quantity} & \colhead{KPNO 0.9m}}
\startdata
Focal Ratio & f/7.5 \nl
Instrument & KPNO 8k x 8k Mosaic \nl
Plate Scale [\arcsec/pixel] & 0.43 \nl
Field Area [$\rm deg^{2}$] & 0.957 \nl
Total Fields               & 171 \nl
Total Area [$\rm deg^{2}$] & 164 \nl
$m_{\rm R50}$\tablenotemark{a} & 21.1 \nl
$\theta$\tablenotemark{b} [\arcsec] & 0.9--2.4 \nl
Filter & $R$ \nl
Quantum Efficiency & 0.85 \nl
\enddata
\tablenotetext{a}{the red magnitude at which detection efficiency reaches half of the maximum efficiency}
\tablenotetext{b}{range of minor axis Full Width at Half Maximum of
stellar sources near focal plane center}
\label{kpnoobs}
\end{deluxetable}
\end{center}

\clearpage

\begin{center}
\begin{deluxetable}{lrcrrrrll}
\small
\tablewidth{0pt}
\tableheadfrac{0.2}
\tablecaption{CFHT Field Centers}
\tablehead{
\colhead{ID} &
\colhead{UT date} &
\colhead{UT times} &
\colhead{$\beta$ \tablenotemark{a}} &
\colhead{$\alpha$ \tablenotemark{b}} &
\colhead{$\delta$ \tablenotemark{c}} &
\colhead{$\theta$ \tablenotemark{d}} &
\colhead{Objects} &
}
\startdata
ecl01.12   & 1998 Sep 25 &  2:54  3:58  4:54 & 12& 22:18:41   &  02:22:01   & p & \nl
ecl01.11   & 1998 Sep 25 &  3:00  4:04  5:00 & 10& 22:20:54   &  00:58:19   & p & \nl
ecl01.09   & 1998 Sep 25 &  3:06  4:10  5:07 &  9& 22:23:06   & -00:25:24   & p & \nl
ecl01.08   & 1998 Sep 25 &  3:13  4:16  5:13 &  8& 22:25:18   & -01:49:07   & p & \nl
ecl01.06   & 1998 Sep 25 &  3:19  4:22  5:19 &  6& 22:27:31   & -03:12:49   & p & \nl
ecl01.05   & 1998 Sep 25 &  3:25  4:29  5:25 &  4& 22:29:44   & -04:36:29   & p & \nl
ecl01.03   & 1998 Sep 25 &  3:31  4:35  5:31 &  3& 22:31:56   & -06:00:00   & m & \nl
ecl01.02   & 1998 Sep 25 &  3:38  4:41  5:38 &  1& 22:34:11   & -07:38:00   & m & \nl
ecl01.00   & 1998 Sep 25 &  3:44  4:48  5:44 &  0& 22:36:26   & -08:47:00   & m & \nl
\hline					         
ecl07.00   & 1998 Sep 25 &  5:50  6:47  7:50 &  0& 23:08:56   & -05:28:00   & p & \nl
ecl07.01   & 1998 Sep 25 &  5:56  6:53  7:56 &  2& 23:06:35   & -04:05:00   & m & \nl
ecl07.03   & 1998 Sep 25 &  6:03  7:00  8:03 &  3& 23:04:17   & -02:42:00   & m & \nl
ecl08.03   & 1998 Sep 25 &  6:09  7:06  8:09 &  3& 23:09:39   & -02:09:00   & p & \nl
ecl08.01   & 1998 Sep 25 &  6:15  7:12  8:16 &  2& 23:11:59   & -03:31:00   & m & \nl
ecl08.00   & 1998 Sep 25 &  6:22  7:25  8:22 &  0& 23:14:20   & -04:54:00   & p & \nl
ecl09.00   & 1998 Sep 25 &  6:29  7:31  8:28 &  0& 23:19:43   & -04:20:00   & p & \nl
ecl09.01   & 1998 Sep 25 &  6:35  7:37  8:35 &  2& 23:17:22   & -02:57:00   & p & \nl
ecl09.03   & 1998 Sep 25 &  6:41  7:44  8:41 &  3& 23:15:01   & -01:34:00   & p & \nl
\hline					         
ecl25.00   & 1998 Sep 25 &  8:50  9:50 10:53 &  0& 00:46:09   &  04:59:51   & p & \nl
ecl25.01   & 1998 Sep 25 &  8:56  9:56 10:59 &  2& 00:43:48   &  06:20:15   & p & \nl
ecl25.03   & 1998 Sep 25 &  9:03 10:10 11:05 &  3& 00:41:26   &  07:43:04   & p & \nl
ecl26.03   & 1998 Sep 25 &  9:10 10:17 11:11 &  3& 00:46:56   &  08:17:41   & p & \nl
ecl26.01   & 1998 Sep 25 &  9:16 10:23 11:17 &  1& 00:49:17   &  06:54:48   & p & \nl
ecl26.00   & 1998 Sep 25 &  9:22 10:29 11:23 &  0& 00:51:38   &  05:31:52   & p & \nl
ecl27.00   & 1998 Sep 25 &  9:28 10:35 11:29 &  0& 00:57:08   &  06:06:12   & p & \nl
ecl27.01   & 1998 Sep 25 &  9:37 10:41 11:35 &  1& 00:54:47   &  07:29:13   & p & \nl
ecl27.03   & 1998 Sep 25 &  9:43 10:47 11:41 &  3& 00:52:26   &  08:52:12   & p & \nl
\hline					         
ecl02.12   & 1998 Sep 26 &  2:36  3:33  4:29 & 12& 22:23:58   &  02:53:28   & m & \nl
ecl02.11   & 1998 Sep 26 &  2:42  3:39  4:37 & 11& 22:26:12   &  01:29:55   & m & \nl
ecl02.09   & 1998 Sep 26 &  2:48  3:45  4:43 &  9& 22:28:25   &  00:06:19   & g & \nl
ecl02.08   & 1998 Sep 26 &  2:54  3:51  4:50 &  8& 22:30:39   & -01:17:15   & g & \nl
ecl02.06   & 1998 Sep 26 &  3:00  3:57  4:57 &  6& 22:32:52   & -02:40:49   & m & \nl
ecl02.05   & 1998 Sep 26 &  3:06  4:04  5:03 &  5& 22:35:07   & -04:04:21   & m & \nl
ecl02.03   & 1998 Sep 26 &  3:13  4:10  5:09 &  3& 22:37:22   & -05:27:55   & g & \nl
ecl02.02   & 1998 Sep 26 &  3:19  4:17  5:15 &  1& 22:39:37   & -06:51:22   & m & \nl
ecl02.00   & 1998 Sep 26 &  3:26  4:23  5:35 &  0& 22:41:52   & -08:14:50   & m & \nl
\hline					         
ecl10.00   & 1998 Sep 26 &  5:42  6:38  7:43 &  0& 23:25:06   & -03:45:45   & m & \nl
ecl10.01   & 1998 Sep 26 &  5:48  6:44  7:49 &  2& 23:22:44   & -02:23:02   & m & \nl
ecl10.03   & 1998 Sep 26 &  5:54  7:00  7:56 &  3& 23:20:23   & -01:00:15   & g & \nl
ecl11.03   & 1998 Sep 26 &  6:00  7:06  8:00 &  3& 23:25:44   & -00:25:49   & g & \nl
ecl11.01   & 1998 Sep 26 &  6:07  7:12  8:08 &  2& 23:28:06   & -01:48:33   & g & 1998 $\rm SN_{165}$ \nl
ecl11.00   & 1998 Sep 26 &  6:13  7:18  8:14 &  0& 23:30:29   & -03:11:16   & g & \nl
ecl12.00   & 1998 Sep 26 &  6:19  7:25  8:20 &  0& 23:35:52   & -02:36:35   & g & \nl
ecl12.01   & 1998 Sep 26 &  6:25  7:31  8:27 &  2& 23:33:29   & -01:13:53   & m & \nl
ecl12.03   & 1998 Sep 26 &  6:31  7:37  8:33 &  3& 23:31:07   &  00:08:46   & m & \nl
\hline					         
ecl28.00   & 1998 Sep 26 &  8:39  9:49 10:50 &  0& 01:02:37   &  06:40:22   & m & \nl
ecl28.01   & 1998 Sep 26 &  8:45  9:56 10:56 &  1& 01:00:18   &  08:03:29   & g & \nl
ecl28.03   & 1998 Sep 26 &  8:59 10:02 11:02 &  3& 00:57:58   &  09:26:32   & m & \nl
ecl29.03   & 1998 Sep 26 &  9:05 10:08 11:08 &  3& 01:03:31   &  10:00:40   & m & \nl
ecl29.01   & 1998 Sep 26 &  9:12 10:14 11:15 &  1& 01:05:50   &  08:37:32   & m & \nl
ecl29.00   & 1998 Sep 26 &  9:18 10:20 11:22 &  0& 01:08:09   &  07:14:19   & m & \nl
ecl30.00   & 1998 Sep 26 &  9:24 10:26 11:28 &  0& 01:13:41   &  07:48:03   & m & \nl
ecl30.01   & 1998 Sep 26 &  9:31 10:32 11:34 &  1& 01:11:24   &  09:11:22   & m & \nl
ecl30.03   & 1998 Sep 26 &  9:37 10:38 11:40 &  3& 01:09:05   &  10:34:40   & g & \nl
ecl31.00   & 1998 Sep 26 &  9:44 10:44 11:46 &  0& 01:19:15   &  08:21:34   & g & \nl
\hline					         
ecl03.06   & 1998 Sep 27 &  2:26  3:26  4:36 &  6& 22:38:15   & -02:08:29   & m & \nl
ecl03.05   & 1998 Sep 27 &  2:32  3:33  4:42 &  4& 22:40:30   & -03:31:54   & m & \nl
ecl03.03   & 1998 Sep 27 &  2:38  3:39  4:48 &  3& 22:42:44   & -04:55:20   & m & \nl
ecl03.02   & 1998 Sep 27 &  2:44  3:51  4:56 &  2& 22:45:01   & -06:18:43   & p & \nl
ecl03.00   & 1998 Sep 27 &  2:50  3:57  5:02 &  0& 22:47:18   & -07:42:04   & m & \nl
ecl04.00   & 1998 Sep 27 &  2:56  4:04  5:08 &  0& 22:52:44   & -07:08:59   & m & \nl
ecl04.02   & 1998 Sep 27 &  3:02  4:10  5:14 &  1& 22:50:26   & -05:45:46   & m & \nl
ecl04.03   & 1998 Sep 27 &  3:08  4:16  5:20 &  3& 22:48:09   & -04:22:30   & m & \nl
ecl04.05   & 1998 Sep 27 &  3:14  4:22  5:26 &  5& 22:45:52   & -02:59:10   & m & \nl
ecl04.06   & 1998 Sep 27 &  3:20  4:30  5:33 &  6& 22:43:37   & -01:35:52   & p & \nl
\hline					         
ecl13.00   & 1998 Sep 27 &  6:22  7:18  8:25 &  0& 23:41:14   & -02:01:50   & p & \nl
ecl13.02   & 1998 Sep 27 &  6:28  7:24  8:31 &  1& 23:38:52   & -00:39:11   & p & \nl
ecl13.03   & 1998 Sep 27 &  6:36  7:30  8:37 &  3& 23:36:29   &  00:43:26   & p & \nl
ecl14.03   & 1998 Sep 27 &  6:41  7:36  8:43 &  3& 23:41:51   &  01:18:16   & p & \nl
ecl14.02   & 1998 Sep 27 &  6:47  7:43  8:49 &  2& 23:44:14   & -00:04:20   & p & \nl
ecl14.00   & 1998 Sep 27 &  6:53  7:49  8:55 &  0& 23:46:37   & -01:26:56   & p & \nl
ecl15.00   & 1998 Sep 27 &  6:59  7:55  9:01 &  0& 23:51:59   & -00:51:59   & p & \nl
ecl15.02   & 1998 Sep 27 &  7:05  8:02  9:07 &  1& 23:49:37   &  00:30:35   & p & \nl
ecl15.03   & 1998 Sep 27 &  7:11  8:09  9:13 &  3& 23:47:14   &  01:53:11   & m & \nl
\hline					         
ecl31.01   & 1998 Sep 27 &  9:28 10:23 11:17 &  1& 01:16:59   &  09:45:00   & p & \nl
ecl31.03   & 1998 Sep 27 &  9:34 10:29 11:23 &  3& 01:14:40   &  11:08:23   & p & \nl
ecl31.04   & 1998 Sep 27 &  9:42 10:35 11:36 &  5& 01:12:21   &  12:31:42   & p & \nl
ecl31.06   & 1998 Sep 27 &  9:48 10:41 11:42 &  6& 01:10:01   &  13:54:55   & p & \nl
ecl32.06   & 1998 Sep 27 &  9:53 10:47 12:13 &  6& 01:15:39   &  14:28:38   & p & \nl
ecl32.05   & 1998 Sep 27 &  9:59 10:53 11:49 &  5& 01:17:59   &  13:05:18   & p & \nl
ecl32.03   & 1998 Sep 27 & 10:05 10:59 11:55 &  3& 01:20:17   &  11:41:53   & p & \nl
ecl32.03   & 1998 Sep 27 & 10:11 11:05 12:01 &  2& 01:22:34   &  10:18:22   & p & \nl
ecl32.00   & 1998 Sep 27 & 10:17 11:11 12:07 &  0& 01:24:50   &  08:54:49   & m & \nl
\hline					         
ecl05.06   & 1998 Sep 28 &  2:21  3:37  4:37 &  6& 22:48:57   & -01:02:56   & m & \nl
ecl05.05   & 1998 Sep 28 &  2:27  3:43  4:43 &  4& 22:51:14   & -02:26:12   & m & \nl
ecl05.03   & 1998 Sep 28 &  2:33  3:49  4:49 &  3& 22:53:32   & -03:49:22   & g & \nl
ecl05.02   & 1998 Sep 28 &  2:46  3:55  4:55 &  1& 22:55:50   & -05:12:33   & g & \nl
ecl05.00   & 1998 Sep 28 &  2:52  4:01  5:01 &  0& 22:58:08   & -06:35:41   & g & \nl
ecl06.00   & 1998 Sep 28 &  2:58  4:07  5:07 &  0& 23:03:33   & -06:02:07   & g & \nl
ecl06.01   & 1998 Sep 28 &  3:04  4:13  5:13 &  2& 23:01:13   & -04:39:05   & g & \nl
ecl06.03   & 1998 Sep 28 &  3:10  4:19  5:19 &  3& 22:58:54   & -03:15:59   & m & \nl
ecl06.04   & 1998 Sep 28 &  3:16  4:25  5:25 &  4& 22:56:36   & -01:52:54   & g & \nl
ecl06.06   & 1998 Sep 28 &  3:22  4:31  5:31 &  6& 22:54:18   & -00:29:46   & g & \nl
\hline					         
e053m01.0  & 1999 Feb 15 & 10:00 10:54 11:50 & -1& 11:34:10   &  01:47:35   & p & \nl
e053\_00.0 & 1999 Feb 15 & 10:07 11:00 11:56 &  0& 11:34:10   &  02:47:35   & p & \nl
e053p01.0  & 1999 Feb 15 & 10:12 11:06 12:02 &  1& 11:34:10   &  03:47:34   & p & \nl
e054p01.0  & 1999 Feb 15 & 10:18 11:12 12:08 &  1& 11:37:49   &  03:23:56   & p & \nl
e054\_00.0 & 1999 Feb 15 & 10:24 11:19 12:14 &  0& 11:37:49   &  02:23:56   & p & \nl
e054m01.0  & 1999 Feb 15 & 10:30 11:26 12:20 & -1& 11:37:50   &  01:23:53   & p & \nl
e055m01.0  & 1999 Feb 15 & 10:36 11:32 12:26 & -1& 11:41:29   &  01:00:13   & p & \nl
e055\_00.0 & 1999 Feb 15 & 10:42 11:38 12:32 &  0& 11:41:29   &  02:00:13   & p & \nl
e055p01.0  & 1999 Feb 15 & 10:48 11:45 12:38 &  1& 11:41:29   &  03:00:13   & p & \nl
\hline					         
e013p02.0  & 1999 Feb 20 &  3:14  4:33  5:45 &  2& 09:00:36   &  19:00:04   & m & \nl
e013p01.0  & 1999 Feb 20 &  3:20  4:40  5:51 &  1& 09:00:36   &  18:00:07   & m & \nl
e013\_00.0 & 1999 Feb 20 &  3:26  4:46  5:57 &  0& 09:00:36   &  17:00:06   & m & \nl
e013m02.0  & 1999 Feb 20 &  3:32  4:52  6:03 & -2& 09:00:36   &  15:00:06   & g & \nl
e016m01.0  & 1999 Feb 20 &  3:37  4:58  6:09 & -1& 09:12:45   &  15:07:18   & m & \nl
e016\_00.0 & 1999 Feb 20 &  3:43  5:04  6:15 &  0& 09:12:45   &  16:07:17   & g & \nl
e016p01.0  & 1999 Feb 20 &  3:49  5:10  6:21 &  1& 09:12:45   &  17:07:18   & m & \nl
e016p02.0  & 1999 Feb 20 &  3:55  5:16  6:27 &  2& 09:12:45   &  18:07:18   & m & \nl
e023p02.0  & 1999 Feb 20 &  4:01  5:22  6:33 &  2& 09:40:39   &  15:54:37   & m & \nl
e023p01.0  & 1999 Feb 20 &  4:07  5:28  6:39 &  1& 09:40:39   &  14:54:36   & g & \nl
e023\_00.0 & 1999 Feb 20 &  4:13  5:33  6:45 &  0& 09:40:39   &  13:54:37   & m & \nl
e023m01.0  & 1999 Feb 20 &  4:19  5:39  6:51 & -1& 09:40:39   &  12:54:35   & g & \nl
\hline					         
e032m01.0  & 1999 Feb 20 &  6:57  8:10  9:21 & -1& 10:15:38   &  09:47:45   & g & 1999 $\rm DE_{9}$ \nl
e032\_00.0 & 1999 Feb 20 &  7:03  8:16  9:27 &  0& 10:15:38   &  10:47:45   & g & \nl
e032p01.0  & 1999 Feb 20 &  7:09  8:22  9:33 &  1& 10:15:38   &  11:47:45   & g & 1999 $\rm DF_{9}$ \nl
e032p02.0  & 1999 Feb 20 &  7:15  8:28  9:39 &  2& 10:15:38   &  12:47:44   & g & \nl
e040m01.0  & 1999 Feb 20 &  7:46  8:58 10:22 & -1& 10:45:58   &  06:50:10   & g & \nl
e040\_00.0 & 1999 Feb 20 &  7:52  9:03 10:28 &  0& 10:45:58   &  07:50:11   & g & \nl
e040p01.0  & 1999 Feb 20 &  7:58  9:09 10:34 &  1& 10:45:58   &  08:50:11   & g & \nl
e040p02.0  & 1999 Feb 20 &  8:04  9:15 10:40 &  2& 10:45:58   &  09:50:11   & g & \nl
\hline					         
e047m01.0  & 1999 Feb 20 & 10:46 11:36 12:12 & -1& 11:12:03   &  04:08:46   & g & \nl
e047\_00.0 & 1999 Feb 20 & 10:52 11:42 12:18 &  0& 11:12:03   &  05:08:45   & m & \nl
e047p01.0  & 1999 Feb 20 & 10:58 11:48 12:25 &  1& 11:12:03   &  06:08:46   & g & \nl
\hline					         
e1002p05   & 2000 Mar 03 &  3:06  4:22  5:31 &  5& 09:24:33.7 &  20:18:02.1 & g & \nl
e1002p04   & 2000 Mar 03 &  3:19  4:28  5:37 &  4& 09:24:33.7 &  19:17:02.0 & g & \nl
e1002p03   & 2000 Mar 03 &  3:25  4:35  5:44 &  3& 09:24:33.7 &  18:16:01.9 & g & \nl
e1002p02   & 2000 Mar 03 &  3:31  4:41  5:50 &  2& 09:24:33.7 &  17:15:01.8 & g & \nl
e1002p01   & 2000 Mar 03 &  3:37  4:47  5:57 &  1& 09:24:33.7 &  16:14:01.6 & g & \nl
e1002\_00  & 2000 Mar 03 &  3:44  4:53  6:03 &  0& 09:24:33.7 &  15:13:01.5 & g & \nl
e1002m01   & 2000 Mar 03 &  3:50  5:00  6:10 & -1& 09:24:33.7 &  14:12:01.4 & g & \nl
e1002m02   & 2000 Mar 03 &  3:57  5:06  6:15 & -2& 09:24:33.7 &  13:11:01.3 & g & \nl
e1002m03   & 2000 Mar 03 &  4:03  5:12  6:22 & -3& 09:24:33.7 &  12:10:01.2 & g & 2000 $\rm EE_{173}$ \nl
e1002m04   & 2000 Mar 03 &  4:09  5:18  6:28 & -4& 09:24:33.7 &  11:09:01.0 & g & \nl
e1002m05   & 2000 Mar 03 &  4:15  5:25  6:34 & -5& 09:24:33.7 &  10:08:00.9 & g & \nl
\hline					         
e1020m05   & 2000 Mar 03 &  6:42  7:51  9:00 & -5& 10:40:33.6 &  03:17:43.3 & g & \nl
e1020m04   & 2000 Mar 03 &  6:48  7:57  9:06 & -4& 10:40:33.6 &  04:18:43.4 & g & \nl
e1020m03   & 2000 Mar 03 &  6:54  8:03  9:13 & -3& 10:40:33.6 &  05:19:43.5 & g & \nl
e1020m02   & 2000 Mar 03 &  7:00  8:10  9:19 & -2& 10:40:33.6 &  06:20:43.6 & g & \nl
e1020m01   & 2000 Mar 03 &  7:07  8:16  9:26 & -1& 10:40:33.6 &  07:21:43.8 & g & \nl
e1020\_00  & 2000 Mar 03 &  7:13  8:22  9:32 &  0& 10:40:33.6 &  08:22:43.9 & g & \nl
e1020p01   & 2000 Mar 03 &  7:19  8:28  9:38 &  1& 10:40:33.6 &  09:23:44.0 & g & \nl
e1020p02   & 2000 Mar 03 &  7:24  8:35  9:45 &  2& 10:40:33.6 &  10:24:44.1 & g & \nl
e1020p03   & 2000 Mar 03 &  7:32  8:41  9:51 &  3& 10:40:33.6 &  11:25:44.2 & g & \nl
e1020p04   & 2000 Mar 03 &  7:38  8:47  9:57 &  4& 10:40:33.6 &  12:26:44.4 & g & \nl
e1020p05   & 2000 Mar 03 &  7:45  8:54 10:03 &  5& 10:40:33.6 &  13:27:44.5 & g & \nl
\hline					         
e1041m05   & 2000 Mar 03 & 10:11 10:56 11:55 & -5& 12:05:04.3 & -05:37:58.9 & m & \nl
e1041m04   & 2000 Mar 03 & 10:17 11:03 12:01 & -4& 12:05:04.3 & -04:36:58.8 & g & \nl
e1041m03   & 2000 Mar 03 & 10:23 11:09 12:08 & -3& 12:05:04.3 & -03:35:58.6 & g & \nl
\hline					         
e1003m05   & 2000 Mar 04 &  2:28  3:44  4:47 & -5& 09:28:54.5 &  09:47:19.9 & m & \nl
e1003m04   & 2000 Mar 04 &  2:34  3:50  4:54 & -4& 09:28:54.5 &  10:48:20.0 & p & \nl
e1003m03   & 2000 Mar 04 &  2:40  3:57  5:06 & -3& 09:28:54.5 &  11:49:20.1 & m & \nl
e1003m01   & 2000 Mar 04 &  2:59  4:03  5:13 & -1& 09:28:54.5 &  13:51:20.4 & m & \nl
e1003\_00  & 2000 Mar 04 &  3:05  4:10  5:20 &  0& 09:28:54.5 &  14:52:20.5 & m & \nl
e1003p01   & 2000 Mar 04 &  3:13  4:16  5:26 &  1& 09:28:54.5 &  15:53:20.6 & m & \nl
e1003p02   & 2000 Mar 04 &  3:19  4:22  5:32 &  2& 09:28:54.5 &  16:54:20.7 & m & \nl
e1003p03   & 2000 Mar 04 &  3:25  4:29  5:38 &  3& 09:28:54.5 &  17:55:20.8 & m & \nl
e1003p04   & 2000 Mar 04 &  3:32  4:35  5:45 &  4& 09:28:54.5 &  18:56:21.0 & m & \nl
e1003p05   & 2000 Mar 04 &  3:38  4:41  5:51 &  5& 09:28:54.5 &  19:57:21.1 & g & \nl
\hline					         
e1021m04   & 2000 Mar 04 &  5:58  6:55  8:14 & -4& 10:44:39.7 &  03:54:03.5 & m & \nl
e1021m03   & 2000 Mar 04 &  6:04  7:01  8:20 & -3& 10:44:39.7 &  04:55:03.6 & m & \nl
e1021m02   & 2000 Mar 04 &  6:10  7:07  8:27 & -2& 10:44:39.7 &  05:56:03.7 & m & \nl
e1021m01   & 2000 Mar 04 &  6:16  7:25  8:33 & -1& 10:44:39.7 &  06:57:03.8 & m & \nl
e1021\_00  & 2000 Mar 04 &  6:23  7:32  8:39 &  0& 10:44:39.7 &  07:58:04.0 & m & \nl
e1021p01   & 2000 Mar 04 &  6:29  7:40  8:45 &  1& 10:44:39.7 &  08:59:04.1 & p & \nl
e1021p02   & 2000 Mar 04 &  6:35  7:46  8:52 &  2& 10:44:39.7 &  10:00:04.2 & p & \nl
e1021p03   & 2000 Mar 04 &  6:41  7:52  8:58 &  3& 10:44:39.7 &  11:01:04.3 & p & \nl
\hline					         
e1035p05   & 2000 Mar 04 &  9:26 10:17 11:10 &  5& 11:41:11.8 &  07:07:06.2 & m & \nl
e1036p05   & 2000 Mar 04 &  9:32 10:23 11:17 &  5& 11:45:11.1 &  06:41:15.9 & m & \nl
e1037p05   & 2000 Mar 04 &  9:39 10:29 11:23 &  5& 11:49:10.1 &  06:15:24.8 & p & \nl
\enddata
\tablecomments{This table lists fields imaged with the KPNO 8k Mosaic
camera.  Fields were imaged in triplets, with UT times given for each
image.  KBOs found are listed after the field of discovery.  The
complete version of this table is in the electronic edition of the
Journal.  The printed edition contains only a sample.}
\tablenotetext{a}{J2000 ecliptic latitude, degrees}
\tablenotetext{b}{J2000 right ascension, hours}
\tablenotetext{c}{J2000 declination, degrees} \tablenotetext{d}{Seeing
category: g, m, p represent the good ($< 1.5$ arc sec), medium ($\geq
1.5$ arc sec and $< 2.0$ arc sec), and poor ($\geq 2.0$ arc sec)
seeing cases, respectively.  The efficiency functions for each of
these cases are presented in Table~\ref{efftable}.}
\label{kpnofieldtable} \end{deluxetable} \end{center}

\clearpage

\begin{center}
\begin{deluxetable}{ccccc}
\small
\tablewidth{0pt}
\tablecaption{KPNO Survey Efficiency}
\tablehead{\colhead{} & \colhead{Good} & \colhead{Medium} & \colhead{Poor} & \colhead{Global}}
\startdata
Median PSF FWHM [\arcsec]       & 1.3      & 1.6      & 2.1      & 1.5      \nl
PSF FWHM Range  [\arcsec]       & 1.0--1.4 & 1.5--1.9 & 2.0--2.4 & 1.0--2.4 \nl
$e_{\rm max}$                   & 0.88     & 0.88     & 0.80     & 0.85     \nl
$m_{R50}$                       & 21.4     & 21.2     & 20.8     & 21.1     \nl
$\sigma$                        & 0.5      & 0.5      & 0.4      & 0.6      \nl
Fields Imaged                   & 57       & 61       & 53       & 171      \nl
Total Area [$\rm deg^{2}$]      & 55       & 58       & 51       & 164      \nl
KBOs                            & 3        & 0        & 0        & 3        \nl
Centaurs                        & 1        & 0        & 0        & 1        \nl
\enddata
\tablecomments{Survey efficiency for three seeing cases and ``Global''
efficiency, representing the sky-area weighted mean of the three
efficiencies, designed to represent the properties of the KPNO data
set in its entirety.}
\label{efftable}
\end{deluxetable}
\end{center}

\clearpage

\begin{center}
\begin{deluxetable}{ccccc}
\small
\tablewidth{0pt}
\tablecaption{The KPNO Objects}
\tablehead{ \colhead{} & \colhead{1998 $\rm SN_{165}$\tablenotemark{a}} & \colhead{1999
$\rm DE_{9}$} & \colhead{1999 $\rm DF_{9}$} & \colhead{2000 $\rm EE_{173}$}}
\startdata
Discovery Conditions \nl
Date                       & 1998 Sep 26         & 1999 Feb 20         & 1999 Feb 20         & 2000 Mar 03           \nl
$R$ [AU]                   & 38.248              & 32.549              & 39.765              & 23.593                \nl
$\Delta$ [AU]              & 37.264              & 33.537              & 38.776              & 22.678                \nl
$\alpha'$ [deg]            & 0.3                 & 0.0                 & 0.0                 & 0.9                   \nl
$m_{\rm R}$                & $21.2 \pm 0.1$      & $19.7 \pm 0.1$      & $21.6 \pm 0.1$      & $22.2 \pm 0.1$        \nl
$m_{\rm R}(1,1,0)$         & 5.4                 & 4.5                 & 5.7                 & 8.4                   \nl
Diameter [km]              & 480                 & 710                 & 420                 & 120                   \nl
$a$ [AU]                   & $37.900 \pm 0.003$  & $56.4 \pm 0.5$      & $46.400 \pm 0.053$  & $49.02 \pm 0.06$      \nl
$e$                        & $0.0437 \pm 0.0004$ & $0.435 \pm 0.009$   & $0.144 \pm 0.004$   & $0.5387 \pm 0.0007$   \nl
$i$ [deg]                  & $4.605 \pm 0.0005$  & $7.623 \pm 0.002$   & $9.819 \pm 0.005$   & $5.950 \pm 0.0004$    \nl
$\Omega$ (node) [deg]      & $192.097 \pm 0.002$ & $322.983 \pm 0.003$ & $334.856 \pm 0.002$ & $294.070 \pm 0.002$   \nl
$\omega$  (peri) [deg]     & $264.156 \pm 0.145$ & $155.7 \pm 1.7$     & $169 \pm 11$        & $235.05 \pm 0.13$     \nl
$\tau$ (time of peri) [JD] & $2474593 \pm 36$    & $2445972 \pm 222$   & $2449415 \pm 2573$  & $2454160 \pm 10$      \nl
epoch                      & 2451071.7           & 2451229.8           & 2451229.8           & 2451606.7             \nl
perihelion                 & 36.33554            & 31.83661            & 39.70701            & 22.62496              \nl
\enddata
\tablecomments{Discovery conditions of the objects found in this
survey.  $R$ is the heliocentric distance. $\Delta$ is geocentric
distance. $\alpha'$ is phase angle.  $m_{\rm R}$ is the red magnitude
at discovery.  $m_{\rm R}(1,1,0)$ is red magnitude at zero phase
angle, with heliocentric and geocentric distance = 1 AU.  The diameter
is derived assuming a geometric albedo of 4\%.  Orbital elements and
associated errors were computed using the procedure of Bernstein and
Khushalani (2000) with data provided by the Minor Planet Center.}
\tablenotetext{a}{This object was discovered by Spacewatch on Sep 23,
1998 and serendipitously detected in our data.}
\label{kpnodisc}
\end{deluxetable}
\end{center}

\clearpage

\begin{center}
\begin{deluxetable}{llll}
\small
\tablewidth{0pt}
\tablecaption{Classical KBO Model Assumptions}
\tablehead{\colhead{Quantity} & \colhead{Value} &
\colhead{Distribution} & \colhead{Description}}
\startdata
$a$ [AU]                                          & $41 < a < 47 $ & $n(a)da \sim a^{1-p} da$ & semimajor axis distribution \nl
$p$\tablenotemark{a}                              & 2                          & ---                      & semimajor axis power \nl
$q'$\tablenotemark{b} [AU]                        & $q' < 37.0 $               & ---                      & perihelion criterion \nl
$i$                                               & ---                        & Gaussian                 & inclination distribution\nl
$i_{1/2}$\tablenotemark{c} [deg]                  & 20.0                       & ---                      & inclination half-width \nl
$r$ [km]                                          & $25 < r < 16000$           & $n(r) dr \sim r^{-q} dr$ & radius distribution \nl
$q$                                               & fitted                     & ---                      & radius distribution power \nl
$N_{\rm CKBOs}(D > 100 \mbox{ km})$               & fitted                     & ---                      & number of CKBOs with $D > 100$ km \nl
$N_{\rm bins}$                                    & 50                         & logarithmic              & number of radius bins \nl
\enddata
\tablenotetext{a}{For circular orbits, $p$ represents the decrease in
ecliptic plane surface density $\Sigma_{\rm ecl}$ as a function of
heliocentric distance $R$, $\Sigma_{\rm ecl} \sim R^{-p}$.}
\tablenotetext{b}{Eccentricity assumed to be uniform between 0 and 1
consistent with perihelion criterion.}
\tablenotetext{c}{Best-fit values from Trujillo, Jewitt \& Luu (2001).}
\label{model}
\end{deluxetable}
\end{center}

\clearpage

\begin{center}
\begin{deluxetable}{ccccc}
\small
\tablewidth{0pt}
\tablecaption{Objects Brighter than 20 R}
\tablehead{ \colhead{} & \colhead{1996 $\rm GQ_{21}$} & \colhead{1999
$\rm DE_{9}$} & \colhead{2000 $\rm EB_{173}$}}
\startdata
Discovery Conditions \nl
Date                       & 1996 Apr 12          & 1999 Feb 20         & 2000 Mar 10          \nl
$R$ [AU]                   & 38.523               & 32.549              & 29.940               \nl
$\Delta$ [AU]              & 37.521               & 33.537              & 29.036               \nl
$\alpha'$ [deg]            & 0.1                  & 0.0                 & 0.8                  \nl
$m$                        & 20.5 V               & 19.7 R              & 19.3 R               \nl
$m(1,1,0)$                 & 4.7 V                & 4.5 R               & 4.5 R                \nl
Diameter [km]              & 720                  & 710                 & 780                  \nl
$a$ [AU]                   & $94.06 \pm 0.10$     & $56.4 \pm 0.5$      & $39.22 \pm 0.03$     \nl
$e$                        & $0.5955 \pm 0.0007$  & $0.435 \pm 0.009$   & $0.2697 \pm 0.0011$  \nl
$i$ [deg]                  & $ 13.357 \pm 0.0003$ & $7.623 \pm 0.002$   & $15.476 \pm 0.0004$  \nl
$\Omega$ (node) [deg]      & $194.184 \pm 0.0003$ & $322.983 \pm 0.003$ & $169.331 \pm 0.001$  \nl
$\omega$  (peri)[deg]      & $353.992 \pm 0.3$    & $155.7 \pm 1.7$     & $66.8 \pm 0.2$       \nl
$\tau$ (time of peri) [JD] & $2447365 \pm 54$     & $2445972 \pm 222$   & $2456944 \pm 22$     \nl
epoch [JD]                 & 2450185.8            & 2451229.8           & 2450182.7            \nl
source                     & Larsen et al. (2001) & This Work           & Ferrin et al. (2001) \nl
\enddata
\tablecomments{Discovery conditions of all KBOs brighter than 20 R
discovered in published surveys (Pluto excepted).  $R$ is the
heliocentric distance. $\Delta$ is geocentric distance. $m$ is the
magnitude at discovery.  $m(1,1,0)$ is magnitude at zero phase angle,
with geocentric and heliocentric distance = 1 AU.  The diameter is
derived assuming a geometric albedo of 4\%.  Orbital elements and
associated errors were computed using the procedure of Bernstein and
Khushalani (2000) with data provided by the Minor Planet Center.}
\label{brightdisc}
\end{deluxetable}
\end{center}

\end{document}